\documentclass[12pt]{article}
\usepackage{graphicx}

\begin{document}

\title{Vertex labeling and routing in expanded Apollonian networks\\
{\normalsize Z. Zhang, F. Comellas, G. Fertin, A. Raspaud, L. Rong, S. Zhou}}
\author{Zhongzhi Zhang\\
{\small Department of Computer Science and Engineering and}\\
{\small Shanghai Key Lab of Intelligent Information Processing,}\\
{\small Fudan University, Shanghai 200433, China}\\
{\small  {\tt zhangzz@fudan.edu.cn}}\\
\vspace{-1mm}\\
Francesc Comellas\\
{\small Dep. de Matem\`atica Aplicada IV,  Universitat Polit\`ecnica de Catalunya}\\
{\small  Av. Canal Ol\'{\i}mpic s/n, 08860
Castelldefels, Barcelona, Catalonia, Spain}\\
{\small  {\tt comellas@ma4.upc.edu}}\\
\vspace{-1mm}\\
Guillaume Fertin\\
{\small  LINA, Universit\'e de Nantes}\\
{\small 2 rue de la Houssini\`ere, BP 92208, 44322 Nantes Cedex 3, France}\\
{\small  {\tt fertin@lina.univ-nantes.fr}}\\
\vspace{-1mm}\\
Andr\'e Raspaud\\
{\small  LaBRI, Universit\'e Bordeaux 1}\\
{\small  351, cours de la Lib\'eration, 33405 Talence Cedex, France}\\
{\small  {\tt raspaud@labri.fr}}\\
\vspace{-1mm}\\
Lili Rong\\
{\small  Institute of Systems Engineering,  Dalian University of Technology}\\
{\small Dalian 116024, Liaoning, China}\\
{\small  {\tt llrong@dlut.edu.cn}}\\
\vspace{-1mm}\\
Shuigeng Zhou\\
{\small Department of Computer Science and Engineering and}\\
{\small Shanghai Key Lab of Intelligent Information Processing,}\\
{\small Fudan University, Shanghai 200433, China}\\
{\small  {\tt  sgzhou@fudan.edu.cn}}
}
\date{}

\maketitle

\begin{abstract}
We present a family of networks, expanded deterministic Apollonian
networks, which are a generalization
of the Apollonian networks  and are
simultaneously scale-free, small-world, and highly clustered. We
introduce a labeling of their vertices that allows to determine a
shortest path routing between any two vertices of the
network based only on the labels.
\end{abstract}

\section{Introduction}
In these last few years there has been a growing interest in the study of
complex networks \cite{AlBa02,DoMe02,Ne03,BoLaMoChHw06}, which can
help to describe many social, biological, and communication
systems, such as co-author networks ~\cite{Ne01}, sexual networks
~\cite{LiEdAmStAb01}, metabolic networks ~\cite{JeToAlOlBa00},
protein networks in the cell ~\cite{JeMaBaOl01},
Internet~\cite{FaFaFa99}, and the World Wide Web~\cite{AlJeBa99}.
Extensive observational studies show that many real-life networks have
at least three important common statistical characteristics: the
degree distribution exhibits a power law tail with an exponent
taking a value between 2 and 3 (\emph{scale-free}); nodes having a
common neighbor are far more likely to be linked to each other than
are two nodes selected randomly (\emph{highly clustered}); the
expected number of links needed to go from one arbitrarily
selected node to another one is low (\emph{small-world property}).

These empirical findings have lead to a new kind  of network
models~\cite{AlBa02,DoMe02,Ne03,BoLaMoChHw06}. The research on
these new  models was started by the two seminal papers by
Watts and Strogatz on small-world networks \cite{WaSt98} and
Barab\'asi and Albert on scale-free networks \cite{BaAl99}.
A wide variety of network models and mechanisms, including initial
attractiveness~\cite{DoMeSa00},  nonlinear preferential
attachment~\cite{KaReLe00}, aging and cost~\cite{AmScBaSt00},
competitive dynamics~\cite{BiBa01}, edge rewiring~\cite{AlBa00} and
removal~\cite{DoMe00b}, duplication~\cite{ChLuDeGa03}, which may
represent processes realistically taking place in real-life systems,
have been proposed.

Recently, based on the classical Apollonian packing, Andrade
\emph{et al.} introduced Apollonian networks~\cite{AnHeAnSi05} which
were simultaneously proposed by Doye and Massen in~\cite{DoMa05}.
Apollonian networks  belong to a deterministic  type of networks
studied earlier in Refs.~\cite{BaRaVi01,CoOzPe00,DoGoMe02,CoSa02,CoFeRa04} which have
received much interest recently~\cite{ZhYaWa05,ZhRoCo05,ZhCoFe05,ZhRoZh06,ZhZh06a}.
Two-dimensional Apollonian networks are simultaneously scale-free, small-world,
Euclidean, space filling, and with matching graphs~\cite{AnHeAnSi05,ZhYaWa05}.
They may provide valuable insight into real-life networks; moreover,
they are maximal planar graphs and this property  is of particular interest
for the layout of printed circuits and related problems~\cite{AnHeAnSi05,ZhYaWa05}.
More recently, some interesting dynamical processes
\cite{ZhYaWa05,ZhRoZh06,HuXuWuWa06}, such as percolation
\cite{ZhYaWa05}, epidemic spreading \cite{ZhYaWa05}, synchronization
\cite{ZhRoZh06}, and random walks \cite{HuXuWuWa06}, taking place on
these networks have been also investigated.

Networks are composed of vertices (nodes) and edges (links)
and are very often studied considering
branch of discrete mathematics known as graph theory.
One active subject in  graph theory is graph
labeling~\cite{Ga05}. This is not only due to its theoretical
importance but also because of the wide range of applications in
many fields~\cite{BlGo77}, such as x-rays, crystallography, coding
theory, radar, astronomy, circuit design, and communication design.

In this paper we present an extension of the  general high dimensional
Apollonian networks~\cite{AnHeAnSi05,DoMa05,ZhCoFe05} which includes
 the deterministic small-world network introduced in~\cite{ZhRoGo05}.
We give a vertex labeling, so that
queries for the shortest path between any two vertices
can be efficiently answered thanks to it. Finding shortest
paths in networks is a well-studied and important problem with also many
applications~\cite{AhMaOr93}. Our  labeling may be useful
in  aspects such as network optimization and information
dissemination, which are directly related to the problem of finding shortest paths
between all pairs of vertices of the network.

\section{Expanded Apollonian networks}
In this section we present a network model defined in a iterative way.
The model, which we call {\em expanded Apollonian network} (EAN),
is an extension of the general high dimensional
Apollonian network~\cite{AnHeAnSi05,DoMa05,ZhCoFe05} which includes
 the deterministic small-world network introduced in~\cite{ZhRoGo05}.

\begin{figure}
\begin{center}
\includegraphics[width=8cm]{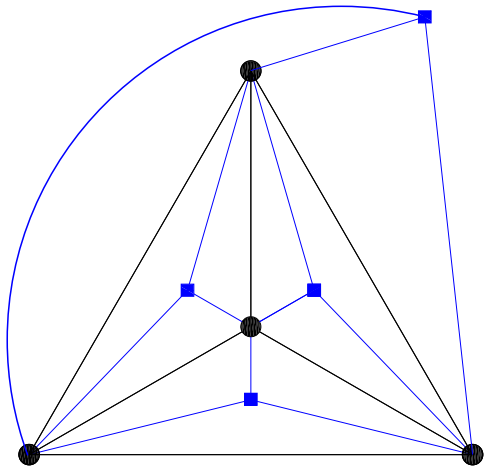}
\caption{Illustration of a growing network in the case of $d=2$,
showing the first two steps of growing process. } \label{network}
\end{center}
\end{figure}

 The networks, denoted by $A(d,t)$ after $t$ iterations with $d\geq 1$
and $t\geq 0$, are constructed as follows. For $t=0$, $A(d,0)$ is a
complete graph $K_{d+2}$ (or $(d+2)$-clique). For $t\geq 1$,
$A(d,t)$ is obtained from $A(d,t-1)$. For each of the existing
subgraphs of $A(d,t-1)$ that is isomorphic to a $(d+1)$-clique and
created at step $t-1$, a new vertex is created and connected to all
the vertices of this subgraph. Figure~\ref{network} shows the
network growing process for the particular case where $d=2$.

Let $n_v(t)$ and $n_e(t)$ denote the number of vertices and edges
created at step $t$, respectively. According to the network
construction, one can see that at step $t_i$ ($t_i>1$) the number of
newly introduced vertices and edges is $n_v(t_i)=(d+2)(d+1)^{t_i-1}$ and
$n_e(t_i)=(d+2)(d+1)^{t_i}$. From these results, we
can easily compute the total number of vertices $N_t$ and edges $E_t$
at step $t$, which are $N_t=\frac{(d+2)[(d+1)^{t}+d-1]}{d}$
and $E_t=(d+2)(d+1) \frac{2(d+1)^{t}+d-2}{2d}$, respectively. So for
large $t$, the average degree $\overline{k}_t= \frac{2E_t}{N_t}$ is
approximately $2(d+1)$.

This general model includes existing models, as listed below.

Indeed, when $d=1$, the network is the deterministic
small-world network (DSWN) introduced in~\cite{ZhRoGo05} and
further generalized in~\cite{ZhRoCo05a}. DSWN is an exponential
network, its degree distribution $P(k)$ is an exponential of a power
of degree $k$. For a node of degree $k$, the exact clustering
coefficient is $\frac{2}{k}$. The average clustering coefficient of
DSWN is $\ln 2$, which approaches to a high constant value 0.6931.
The average path length of DSWN grows logarithmically with the
number of network vertices~\cite{ZhRoZhWa06}.

When $d\geq2$, the networks are exactly the same as the
high-dimensional Apollonian networks (HDAN) with $d$ indicating the
dimension~\cite{AnHeAnSi05,DoMa05,ZhCoFe05,ZhZh06a}. HDAN present
the typical characteristics of real-life networks in nature and
society, and their main topological properties are controlled by
dimension $d$. They have a power law degree distribution with
exponent $\gamma=1+\frac{\ln(d+1)}{\ln d}$ belonging to the interval
between 2 and 3~\cite{AnHeAnSi05,DoMa05,ZhCoFe05,ZhZh06a}. For any
individual vertex in HDAN, its clustering coefficient $C(k)$ is also
inversely proportional to its degree $k$ as
$C(k)=\frac{2d(k-\frac{d+1}{2})}{k(k-1)}$. The mean value $C$ of
clustering coefficient of all vertices in HDAN is very large and is
an increasing function of $d$. For instance, in the special cases
where $d = 2$ and $d = 3$, $C$ asymptotically reaches values
0.8284 and 0.8852, respectively. In addition, HDAN are small worlds.
The diameter of HDAN, defined as the longest shortest distance
between all pairs of vertices, increases logarithmically with the
number of vertices.
So, the EAN model exhibits a transition from an exponential network ($d=1$) to
scale-free networks ($d\geq2$).

\section{Vertex labeling}

Vertex labeling of a network is an assignment of labels to all the
vertices in the network. In most applications, labels are nonnegative integers,
 though in general real numbers could be
used~\cite{Ga05}. In this section, we describe a way to label the
vertices of $A(d,t)$, for any $d\geq 1$ and $t\geq 0$, such that a
routing by shortest paths between any two vertices of $A(d,t)$ can
be deduced from the labels. We note that a more general result on
shortest paths routing of graphs with given treewidth is given
in~\cite{ChZa00}. However, here we address the more specific case
of the expanded Apollonian networks $A(d,t)$. In what follows, we
will denote $L(v)$ as the label of vertex $v$, for any vertex $v$
belonging to $A(d,t)$.

Here the labeling idea, inspired from~\cite{CoFeRa03}, is to assign
to any vertex $v$ created at step $t\geq 1$ a label of length $t$,
in the form of a word of $t$ digits, each digit being an integer
between 1 and $d+2$ (the vertices obtained at step $t=0$, i.e. the
vertices of the {\em initial} ($d+2$)-clique $A(d,0)$, are assigned a
special label). More precisely, the labeling of any vertex $v$ of
$A(d,t)$ is done thanks to the following rules:
\begin{itemize}
\item Label the vertices of the initial ($d+2$)-clique $A(d,0)$
arbitrarily, with labels $1',2'\ldots (d+2)'$.

\item At any step $t\geq 1$, when a new vertex $v$ is added and
joined to all vertices of a clique $K_{d+1}$:

\begin{enumerate}
\item If $v$ is connected to $d+1$ vertices of the initial
($d+2$)-clique, then $L(v)=l$, where $l'$ is the only vertex of the
initial ($d+2$)-clique that does not belong to this $(d+1)$-clique.

\item If not, then $v$ is connected to $w_1,w_2\ldots w_{d+1}$,
where at least one of the $w_i$'s is not a vertex of the initial
($d+2$)-clique. Thus, any such vertex has a label
$L(w_i)=s_{1,i}s_{2,i}\ldots s_{k,i}$. W.l.o.g., let $w_1$ be the
vertex not belonging to the initial ($d+2$)-clique with the longest
label. In that case, we give vertex $v$ the label $L(v)$ defined as
follows: $L(v)=\alpha\cdot L(w_1)$, where $1\leq \alpha\leq d+2$ is
the only integer not appearing as first digit in the labels of
$w_1,w_2\ldots w_{d+1}$, that is $\alpha=\{ 1,2\ldots
d,d+1,d+2\}\slash\cup_{i=1}^{d+1} s_{1,i}$ (the fact that $\alpha$
is unique will be proved by Property 1 below).
\end{enumerate}
\end{itemize}

\begin{figure}
\begin{center}
\includegraphics[width=9cm]{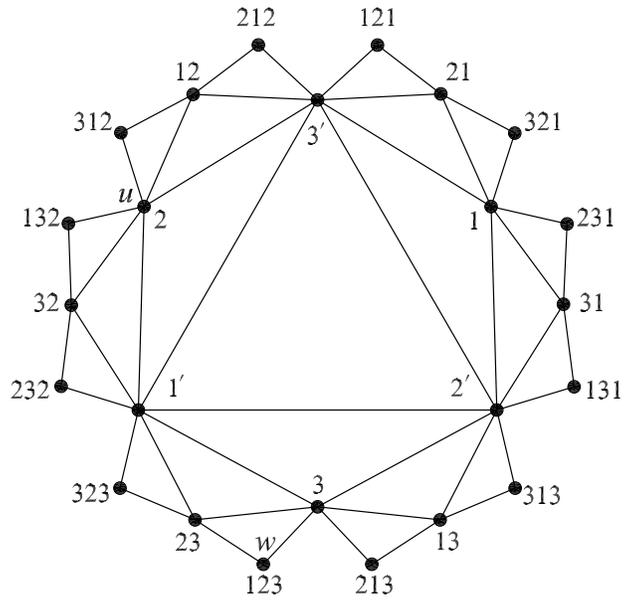}
\includegraphics[width=13cm]{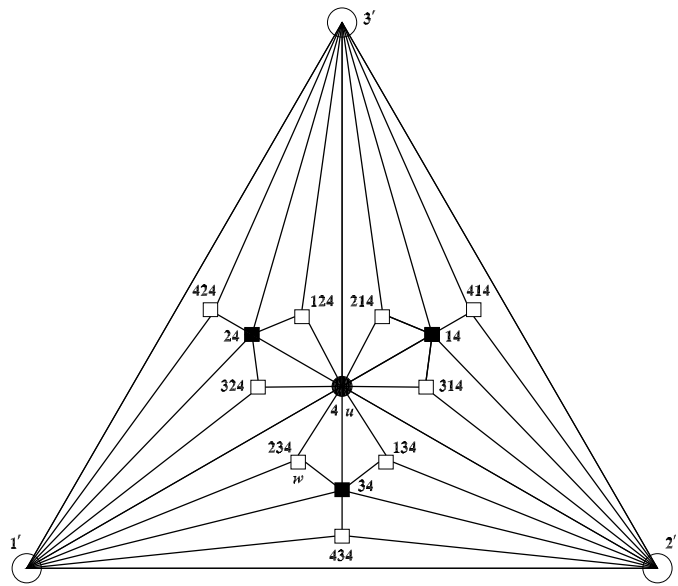}
\caption{(Above) Labels of all vertices of $A(1,3)$. (Below) Labels
of a part of the vertices of $A(2,3)$} \label{labeling01}.
\end{center}
\end{figure}

Such a labeling is illustrated in Figure~\ref{labeling01}. In the
upper part of this figure, we label the vertices of $A(d,t)$, for $d=1$
and up to $t=3$. We see that vertex $u$, created at step 1, has
label $L(u)=2$ because it is not connected to vertex $2'$ of the
initial 3-clique (triangle). Vertex $w$ is is not connected to any
vertex of the initial 3-clique, its label is first composed of the
only digit not appearing as first digit of its neighbors (in this
case, 1), concatenated with the longest label of its neighbors (in
this case, 23). Analogously, in the lower part of
Figure~\ref{labeling01} , where, for sake of clarity, only a part of
$A(2,3)$ is drawn, the vertices have been labeled. For the same
reasons, we can see that vertex $u$ has label $L(u)=4$, while $w$ has
label $L(w)=234$.

\indent Thus, we see that for any $t\geq 1$, any vertex $v_t$
created at step $t$ has a unique label, and that for any vertex $v$
created at step $t\geq 1$, $L(v)=s_1s_2\ldots s_{t}$ is of length
$t$, where each digit $s_j$ satisfies $1\leq s_j\leq d+2$~; while the
vertices created at step $0$ have length 1 (these are the $l'$,
$1\leq l\leq d+2$).

We note that since for any step $t\geq 1$, the number of vertices
that are added to the expanded Apollonian networks is equal to
$(d+2)(d+1)^{t-1}$, the labeling we propose is optimal in the sense
that each label $L(v_t)$ of a vertex created at step $t$ is a
$(d+2)$-ary word of length $t$. Globally, any vertex of $A(d,t)$ is
assigned a label of length $O(\log_{d+2}t)$~; since there are
$N_t=(d+2)\frac {(d+1)^{t}+d-1}{d}$ vertices in $A(d,t)$, we can see
that, overall, the labeling is optimal as well.

\indent Next, we give three properties about the above labeling.
Property 1 ensures that our labeling is deterministic. Property 2 is
a tool to prove Property 3, the latter being important to show that
our routing protocol is valid and of shortest paths.

\textbf{Property 1} In $A(d,t)$, for any ($d+2$)-clique induced by
vertices $w_1$, $w_2$ $\ldots$ $w_{d+2}$, every integer $1\leq i\leq d+2$
appears exactly once as the first digit of the label of a $w_j$.

\textbf{Proof.} By induction on $t$. When $t=1$, the property is
true by construction. Suppose now that the property is true for any
$t'< t$, and let us then show it is true for $t$. Any ($d+2$)-clique
in $A(d,t)$ is composed of exactly one vertex $v$ created at a given
step $t_1$, and $d+1$ vertices $w_1,w_2,\ldots w_{d+1}$ created at
steps strictly less than $t_1$. If $t_1<t$, then the property is
true by induction hypothesis. If $t_1=t$, we suppose that $w_{d+2}$
is connected to a $(d+1)$-clique $\mathcal{C}$ composed of
$w_1,w_2,\ldots w_{d+1}$. It is clear that $\mathcal{C}$ did not
exist at step $t-1$. In other words, one of the $w_i$'s, say $w_1$,
has been created at step $t-1$, based on $d+1$ vertices
$w_2,w_3\ldots w_{d+1}$ and $x$. By induction hypothesis, each
integer $1\leq i\leq d+2$ appears exactly once as first digit of the
labels of $w_1,w_2,w_3\ldots w_{d+1},x$. However, by construction,
the first digit of $L(w_{d+2})$ is the first digit of $L(x)$. Thus
we conclude that each integer $1\leq i\leq d+2$ also appears exactly
once as first digit of the labels of $w_1,w_2,w_3\ldots
w_{d+1},w_{d+2}$, and the result is proved by induction.

\textbf{Property 2} Let $v_t$ be a vertex of $A(d,t)$ created at
step $t\geq 1$. Among the vertices $w_1,w_2,\ldots w_{d+1}$ forming
the $(d+1)$-clique that generated $v_t$, let $w_1,w_2\ldots w_k$,
$k\leq d+1$, be the vertices that do not belong to the initial
($d+2$)-clique. Then $L(v_t)$ is a superstring of $L(w_i)$ for all
$1\leq i\leq k$.

\textbf{Proof.} By induction on $t$. When $t=1$, any vertex $v_1$
created at step 1 is connected to vertices of the initial
($d+2$)-clique only. Thus the result is true. Now suppose the result
is true for any $1\leq t'\leq t-1$, $t\geq 2$, and let us prove it
is then true for $t$. For this, we consider a vertex $v_t$ created
at step $t$, and the $(d+1)$-clique $\mathcal{C}$ it is connected
to. Suppose $v_t$ is a neighbor of $w_p$ which was created at step
$t-1$. However, $w_p$ was created itself thanks to a $(d+1)$-clique,
say $\mathcal{C'}$, composed of vertices $x_1,x_2\ldots x_{d+1}$.
W.l.o.g., suppose that $k\leq d+1$ such vertices, $x_1,x_2\ldots
x_k$ do not belong to the initial ($d+2$)-clique. By induction
hypothesis, $L(x_i)\subseteq L(w_p)$ for any $1\leq i\leq k$. Hence,
in $\mathcal{C}$, $w_p$ is the vertex not belonging to the initial
($d+2$)-clique that has the longest label. By construction of
$L(v_t)$, we have that $L(w_p)\subseteq L(v_t)$, thus we also
conclude that $L(x_i)\subseteq L(v_t)$ for any $1\leq i\leq k$. Thus
$L(v_t)$ is a superstring of the labels of any vertex of
$\mathcal{C}$ that does not belong to the initial ($d+2$)-clique,
and the result is proved by induction.


\textbf{Property 3 } Let $v_t$ be a vertex of $A(d,t)$ created at
step $t\geq 1$. For any $1\leq i\leq d+2$, if $i\not\in L(v_t)$,
then $v_t$ is a neighbor of a vertex $v'$ of the initial
($d+2$)-clique, such that $L(v')=i'$.

\textbf{Proof.} By induction on $t$. When $t=1$, any vertex $v_1$
constructed at step 1 is assigned label $i$, where $i'$ is the only
vertex of the initial ($d+2$)-clique $v_t$ is not connected
to~; thus, by construction, the property is satisfied.\\
\indent Now we suppose that the property is true for any $1\leq
t'\leq t-1$, $t\geq 2$, and we will show it then holds for $t$ as
well. As for the previous property, we consider a vertex $v_t$
created at step $t$, and the $(d+1)$-clique $\mathcal{C}$ it is
connected
to.\\
\indent Suppose $v_t$ is connected to a vertex $w_{t-1}$ that was
created at step $t-1$. However, $w_{t-1}$ was created itself thanks
to a $(d+1)$-clique $\mathcal{C'}$ composed of vertices
$x_1,x_2\ldots x_{d+1}$. Among those $d+1$ vertices, only one, say
$x_p$, does not belong to $\mathcal{C}$. W.l.o.g., suppose that
$k\leq d+1$ such vertices, $x_1,x_2\ldots x_k$ do not belong to the
initial ($d+2$)-clique. Now suppose that $i\not\in L(v_t)$~; then $i$
appears as the first digit of one of the $L(x_j)$'s,
$j\in[1,p-1]\bigcup [p+1,d+1]$, or of $L(w_{t-1})$ (by Property 1).
However, $L(x_j)\subseteq L(w_{t-1})\subseteq L(v_t)$ for any $1\leq
j\leq k$ (by Property 2). Thus, neither $w_{t-1}$ nor any vertex
among the $x_j$'s, $1\leq j\leq k$ contains the digit $i$ in its
label. Hence, only a vertex $y$ from the initial ($d+2$)-clique can
have $i$ in its label, and thus $L(y)=i'$. Hence it suffices to show
that $v_t$ and $y$ are neighbors to prove the property. The only
case for which this would not happen is when $y=x_p$~; we will show
that this is not possible. Indeed, by construction of the labels,
the first digit of $L(v_t)$ is the only integer not appearing as
first digit of the labels of the vertices of $\mathcal{C}$, that is
$w_{t-1},x_1,x_2\ldots x_{p-1},x_{p+1}\ldots x_{d+1}$. However, the
fact that we suppose $y=x_p$ means that no vertex of $\mathcal{C}$
contains $i$ in its label. Thus this would mean that the first digit
of $L(v_t)$ is $i$, a contradiction. Thus, $v_t$ is connected to $y$
with $L(y)=i'$, and the induction is proved.


\section{Routing by shortest path} \label{sec:routing}

Now we describe the routing protocol between any two vertices $u$
and $v$ of $A(d,t)$, with labels respectively equal to $L(u)$ and
$L(v)$. We note that since $A(d,0)$ is isomorphic to the complete
graph $K_{d+2}$, we can assume $t\geq 1$. The routing protocol is
special here in the sense that the routing is done both from $u$ and
$v$, until they reach a common vertex. Hence, the routing strategy
will be used simultaneously from $u$ and from $v$. In order to find
a shortest path between any two vertices $u$ and $v$, the routing
protocol is as follows. First we compute the longest common suffix
$LCS(L(u),L(v))$ of $L(u)$ and $L(v)$, then we distinguish two
cases:

\begin{enumerate}

\item If $LCS(L(u),L(v))= \emptyset$:

\begin{enumerate}
\item Simultaneously from $u$ and $v$ (say, from $u$): let $u=u_0$
and go from $u_i$ to $u_{i+1}$, $i\geq 0$ where $u_{i+1}$ is the
neighbor of $u_i$ with shortest label.

\item Stop when $u_k$ is a neighbor of the initial ($d+2$)-clique.\\
Let $\bar{L}(u_k)$  (resp. $\bar{L}(v_{k'})$) be the integers not
present in $L(u_k)$ (resp. $L(v_{k'})$), and let $S=\bar{L}(u_k)\cap
\bar{L}(v_{k'})$.

\begin{enumerate}
\item If $S\neq \emptyset$, pick any $l\in S$, and close the path
by taking the edge from $u_k$ to $l'$, and the edge from $l'$ to
$v_{k'}$.

\item If $S=\emptyset$, route from $u_k$ to any neighbor $l'_1$
(belonging to the initial ($d+2$)-clique) of $u_k$, and do
similarly from $v_{k'}$ to a neighbor $l'_2$ (belonging to the
initial ($d+2$)-clique) of $v_{k'}$. Then, take the edge from
$l'_1$ to $l'_2$ and thus close the path from $u$ to $v$.
\end{enumerate}
\end{enumerate}

\item If $LCS(L(u),L(v))\neq \emptyset$, then let us call {\em least
common clique} of $u$ and $v$, or $LCC(u,v)$, the $(d+2)$-clique
composed of the vertex with label $LCS(L(u),L(v))$ and the $d+1$
vertices forming the $(d+1)$-clique that generated the vertex of
label $LCS(L(u),L(v))$. We simultaneously route from $u$ and $v$ to
(respectively) $u_k$ and $v_{k'}$, going each time to the neighbor
with $LCS(L(u),L(v))$ as label suffix, and having the shortest
label. Similarly as above, we stop at $u_k$ (resp. $v_{k'}$), where
$u_k$ (resp. $v_{k'}$) is the first of the $u_i$'s (resp. of the
$v_j$'s) to be a neighbor of $LCC(u,v)$. Then there are two
subcases, depending on $Q={L}(u_k)\cap {L}(v_{k'})$.

\begin{enumerate}
\item If $Q\neq \emptyset$, close the path by going to any
vertex $w$ with label $l$, $l\in Q$.

\item If $Q= \emptyset$, then route from $u_k$ (resp. $v_{k'}$)
to any neighbor $w_1$ (resp. $w_2$) in $LCC(u,v)$, and close the
path by taking the edge $(w_1,w_2)$, which exists since both
vertices $w_1$ and $w_2$ belong to the same clique $LCC(u,v)$.
\end{enumerate}
\end{enumerate}

\textbf{Proposition 1} The above mentioned routing algorithm is valid, and of shortest paths.

\textbf{Proof.} Let us first give the main ideas for the validity of the above
routing protocol. Take any two vertices $u$ and $v$. By construction
of $L(u)$ and $L(v)$, the longest common suffix $LCS(L(u),L(v))$
indicates to which ($d+2$)-clique $u$ and $v$ have to go. We can
consider this as a way for $u$ and $v$ to reach their least common
ancestor in the graph of cliques induced by the construction of
$A(d,t)$, or the ``{\em least common clique}''. In Case~(i), this least
common clique is the initial ($d+2$)-clique~; thus, $u$ and $v$ have
to get back to it. In Case~(ii), the shortest path does not go through
the initial ($d+2$)-clique, and the least common clique of $u$ and
$v$, say $LCC(u,v)$, is indicated by the longest common suffix
$LCS(L(u),L(v))$. In other words, the length of $LCS(L(u),L(v))$
indicates the depth of $LCC(u,v)$ in the graph of cliques induced by
the construction of $A(d,t)$. In that case, the routing is similar
as in Case~(i), except that the initial ($d+2$)-clique has to be
replaced by the clique $LCC(u,v)$. Hence, the idea is to adopt the
same kind of routing, considering only neighbors which also have
$LCS(L(u),L(v))$ as suffix
in their labels.\\
\indent When this least common ancestor is determined, one can see,
still by construction, that the shortest route to reach this clique
(either from $u$ or $v$) is to go to the neighbor which has smallest
label, since the length of the label indicates at which step the
vertex was created. Indeed, the earlier the neighbor $w$
was created, the smaller the distance from $w$ to the least common clique is.\\
\indent After we have reached, from $u$ (resp. from $v$), a vertex
$u_k$ (resp. $v_{k'}$) that is a neighbor of the least common
clique, the last thing we need to know is whether $u_k$ and $v_{k'}$
are neighbors. Thanks to Property 3, we know that looking at
$L(u_k)$ and $L(v_{k'})$ is sufficient to answer this question. More
precisely:

\begin{itemize}
\item In Case~(i)(b)-1, $u_k$ and $v_{k'}$ share a neighbor in the
initial ($d+2$)-clique (by Property 3). All those common neighbors
have label $l'$, where $l\in S$. Hence, if we pick any $l\in S$,
then there exists and edge between $u_k$ and $l'$, as well as an
edge between $l'$ and $v_{k'}$.

\item In Case~(i)(b)-2, $u_k$ and $v_{k'}$ do not share a neighbor
in the initial ($d+2$)-clique. Hence, taking a route from $u_k$
(resp. $v_{k'}$) to any neighbor $l'_1$ (resp. $l'_2$) belonging to
the initial ($d+2$)-clique, we can finally take the edge from $l'_1$
to $l'_2$ (which are neighbors, since they both belong to the
initial ($d+2$)-clique) in order to close the path from $u$ to $v$.

\item In Case~(ii)(a), $u_k$ and $v_{k'}$ share a neighbor in
$LCC(u,v)$. Hence we can close the path by going to any vertex $w$
with label $l$, $l\in Q$, since $w$ is a neighbor of both $u_k$ and
$v_{k'}$.

\item In Case~(ii)(b), $u_k$ and $v_{k'}$ do not share a neighbor in
$LCC(u,v)$. Hence we route from $u_k$ (resp. $v_{k'}$) to any
neighbor $w_1$ (resp. $w_2$) in $LCC(u,v)$, and we close the path by
taking the edge $(w_1,w_2)$. This edge exists since both vertices
$w_1$ and $w_2$ belong to the same clique $LCC(u,v)$.
\end{itemize}

\indent Hence we conclude that our labeling of vertices in $A(d,t)$
allows a routing between any two vertices $u$ and $v$, and that it
is of shortest paths.

\section{Conclusion}

We have proposed an expanded deterministic Apollonian
network model, which represents a transition for degree distribution
between exponential and power law distributions. Our model
successfully reproduces some remarkable characteristics in many
nature and man-made networks. We have also introduced a  vertex  labeling
for  these networks. The length of the label is optimal.
Using the vertex labels it is possible to find in an efficient way a
shortest path between any pair of vertices.
Nowadays, efficient handling and delivery in communication
networks (e.g. the Internet) has become one  important
practical issues, and it is directly related to  the problem of finding shortest
paths between any two vertices. Our results, therefore,  can be useful
when describing new communication protocols for complex communication systems.

\subsection*{Acknowledgment}
Zz.Z. and Sg.Z. gratefully acknowledge partial support from the
National Natural Science Foundation of China under Grant Nos.
60373019, 60573183, and 90612007. Support for F.C. was provided by
the Secretaria de Estado de Universidades e Investigaci\'on
(Ministerio de Educaci\'on y Ciencia),  Spain, and the European
Regional Development Fund (ERDF) under project TEC2005-03575.



\end{document}